\documentstyle[aps]{revtex}
\begin{document}
\def\beq{\begin{eqnarray}}
\def\eeq{\end{eqnarray}}
\title{The Error in Bell's Theorem}
\draft
\author{A. F. Kracklauer}
\address{Belvederer Allee 23c \\Weimar, Germany\\
kracklau@fossi.uni-weimar.de}
\date{\today}
\twocolumn[           
\maketitle
\begin{minipage}{\textwidth}   
\begin{quotation}           

\begin{abstract}
An error in the proof of Bell's Theorem is identified and a 
semiclassical model of the EPRB experiment is presented.
\bigskip   
\end{abstract}
\pacs{3.65.Bz, 3.65.Sq}
\end{quotation}      
\end{minipage}]      

In Quantum Mechanics (QM) Bell's Theorem has ensconced as dogma a 
concept, nonlocality, that some researchers consider profoundly 
antirational. \cite{Bell} Further, because this nonlocal `effect' has 
not been correlated with any of the four forces which constitute the 
substance of physics, its ontological status is undetermined. 
\cite{AB} Finally, nonlocality is strikingly absent from the 
elucidation of worldly effects; it makes its appearance as the 
rationalization for no explanation.  In spite of all this, it remains 
an almost beloved special tenet of QM. 

The point of this letter, in this context, is to show that Bell's 
theorem contains an error invalidating its proof.  \cite{KSB} 

To begin and to set notation, I review the basics of QM pertinent to 
the `photon' variant of the Einstein-Podolsky-Rosen-Bohm  (EPRB) 
experiment upon which proofs of Bell's Theorem are based. Consider a 
singlet state function for a pair of photons 

\beq
\psi = {1\over{\sqrt{2}}}\left[ \vert x_1 \rangle\vert y_2 \rangle-
          \vert x_2 \rangle \vert y_1 \rangle \right], 
\label{singlet}\eeq
where $x_1, y_1$, etc. indicate photons polarized in the $x, y$ 
direction respectively and propagating away from each other along the 
$z$ direction.  The photons are then, in one variant, analyzed 
(measured) by passage through, say, Wollaston prisms which separate 
the polarization states spatially; i.e, they perform dichotomic 
measurements, as a photon can be found in one of the two exit 
channels, labeled $+1$ or $-1$. 

The calculation of the state function of the system when one 
channel is rotated by $\phi$ about the axial direction with respect to 
the other, is obtained from  Eq.~\ref{singlet}, using the standard 
rotation matrix: \beq \left[\matrix{\vert x_1'\rangle \cr \vert y_1' 
\rangle}\right]= \left[ \matrix{\cos(\phi) & \sin(\phi) \cr                  
              -\sin(\phi) & \cos(\phi)}  \right]
\left[ \matrix{\vert x_1\rangle \cr \vert y_1 \rangle} \right].
\label{Xform}\eeq
The result is:
\beq
\psi' & = & {1\over{\sqrt{2}}} \left[\right. \cos(\phi)\vert x_1
\rangle\vert y_2' \rangle  -\cos(\phi)\vert y_1 \rangle\vert 
x_2' \rangle   \nonumber \\ 
&    &  +\sin(\phi)\vert x_1 \rangle\vert x_2' \rangle +
  \sin(\phi)\vert y_1 \rangle\vert y_2' \rangle 
\left.\right]. 
\eeq 

From this expression, the essential QM results follow 
directly.  The correlation coefficient is computed as follows:
\beq
\psi'^*M_1 M_2 \psi' = -\cos(2\phi),
\eeq
where $M_1$ is a `measurement' made on photon 1; i.e., $M_1 \vert x_1 
\rangle=+1 \vert x_1 \rangle,\,\, M_1 \vert y_1\rangle=-1 \vert 
y_1\rangle$.   The probability  that both photons are polarized in the 
$x$-directions respectively, for example, is the square of the third 
term: 
\beq 
P(x_1',x_2')={1\over{2}}\sin^2(\phi).
\eeq

The salient point here for present purposes is that a `state;' 
e.g., $\vert x_1 \rangle$, transforms per Eq.~\ref{Xform}.  This 
means, specifically, that viewed just as a function of $\phi$, $\vert 
x_1 \rangle$ takes the form: 
\beq 
\vert x_j' \rangle = \cos(\phi) +\sin(\phi),
\eeq
an expression that reaches a maximum of $\sqrt{2}$ for various 
values of $\phi$.  

In contrast, in his Theorem, Bell considered functions (which 
correspond to a state---or signal---squared) labeled $A$ and $B$ which 
he required to be such that $A(\phi)A(\phi) \le 1; \,\,\forall 
{\phi}$.  That is, Bell's functions transform differently under 
changes of ${\phi}$ than those used in QM.  They belong to a different 
and inappropriate class.  Moreover, his choice is not exhaustive; 
there are other objective local models of the EPRB experiment.

Recall that the strategy of `hidden variable' formulations is to 
invest {\it additional variables} into QM functions in order to 
account classically for their well known peculiarities.  Thus Bell, in 
his notation, formally invested  additional variables labeled 
$\lambda$ into $A$ and $B$ to compliment the environmental variables 
already there, ${\bf a}$ and ${\bf b}$, (to include $\phi$ from 
above), and which for the EPRB experiment are polarizer orientations.  
Of course, he also formally enforced locality by requiring $A({\bf 
a},\lambda)$ to be independent of ${\bf b}$ and vice versa.  In any 
case, under transformations of ${\bf a, b}$, in order to conform to 
the `hidden variable' program, $A({\bf a},\lambda)$ and  $B({\bf 
b},\lambda)$ must retain their transformtion properties.  The 
functions introduced by Bell, in spite of very plausible motivation, 
do not qualify.  Below, I suggest functions, ${\cal A, B}$, with 
alternate but still objective local motivation, that do qualify. In 
physical terms, the essential difference is that between a signal 
comprising just one polarization mode (Bell's choice) and one with 
both modes (my choice). 

The main consequence of correcting this error is that Bell 
inequalities, in which the limit is raised above $2\sqrt{2}$,  are no 
longer violated by QM.  Bell's derivation of these inequalities 
explicitly uses the condition that the functions $A, B \le 1$.  When 
this is changed to ${\cal A, B} \le {2}$, new limits are derived which 
the formulas of QM do not exceed.  Arguments for the presence of 
nonlocality in QM are rendered invalid. 

As the presentations in the literature of the derivations of Bell  
inequalities discuss motivation using terminology for countable 
objects (photons) where uncountable radiation is the topic,  
unraveling all the lexical subtleties is tedious.  The central issue, 
however, does not turn on factors introduced by inefficient detectors 
or other geometric compensations.  Analysis of the ideal case demands 
no more than idealized detectors; i.e., assuming that the detector 
count, a stochastic variable $C(t),$  (photoelectric ejected 
electrons) is, for suitably long times $t$, proportional to: 

\beq C(t) 
\propto \int_0^t \int_{V} I({\bf x},t)d^{3}xdt, 
\label{count}\eeq 
where $V$ is the volume of the detector and  $I({\bf x},t)$ is the 
intensity of impinging radiation.  Likewise, special care must be 
taken to adjust for the countable/uncountable problem when converting 
intensities to probabilities in order to identify correctly the 
`normalization' or total count rates. 

Bell's choice of functions to describe classically the EPRB experiment 
clearly was (mis)motivated by seeking to fit presumed dichotomic 
measurement results. \cite{spin} In part the error arises here in 
failing to consider that the wave function state space for 
polarization, $\vert x\rangle \vert y \rangle$, is a Cartesian sum of 
two sets and not a 2-dimensional metric space.  As a consequence, the 
addition rule for elements (in this context this rule is the 
mathematical model of a polarization measurement) is simply a scalar 
rather than a Pythagorean sum. 

An appropriate model using functions with the correct transformation 
properties would be the following:  Consider a cloud of independent 
point sources (e.g., atoms) which emit spherical radiation into $4\pi$ 
steradians (in their own rest frames) with random phase offsets, 
$\gamma$ determined by random locations in the cloud, random motion 
and so on.  And, assume that although a two stage cascade emission is 
facilitated by an intermediate step, that the total radiation from 
each stage alone is unpolarized.  This means that the radiation from 
the source received in each detector will comprise two independent 
random contributions, one each from each polarization state.   

The phase difference between each polarization state then is a random 
function of time.  The signal received at a detector aligned with 
the coordinate system is proportional to the sum of two terms, one for 
each polarization state with its own phase offset, $\gamma_{x,y}$: 
 
\beq 
{\sqrt{\cal A}(\theta, \gamma_x, \gamma_y)} 
=\cos(\theta)\cos(\gamma_x) + \sin(\theta)\cos(\gamma_y), 
\eeq 
and likewise, the signal at a detector oriented at angle $\phi$ to 
the other detector is proportional to: 
\beq 
\lefteqn{{\sqrt{\cal B}(\theta, \phi, \gamma_x, \gamma_y)} = } 
\nonumber \\  & & \cos(\theta + \phi)\cos(\gamma_x) + \sin(\theta + 
\phi)\cos(\gamma_y). 
\eeq

The coincidence count registered by ideal detectors in accord with 
Eq.~\ref{count} will be then proportional to the average over all 
incidence angles, $\theta$, of the squared product of these two 
factors averaged over the phase offsets, $\gamma_{x,y}$: 
 
\beq
{1\over{\pi}}\!\!\int_0^{\pi}\!\![{1\over{\pi^2}}\!\! 
\int_0^{\pi}\!\!\!\!\int_0^{\pi}\!\!\!\!\!\!{\sqrt{\cal A}(\theta, 
\gamma_x, \gamma_y)}{\sqrt{\cal B}(\theta, \phi, \gamma_x, 
\gamma_y)}d\gamma_x d\gamma_y ]^2 d\theta. 
\eeq 

This integral, divided by the ideal {\it pair} production rate (which 
for ideal detectors is simply $1/2$ the single count rate),  yields: 
\beq 
{1\over{2}}\sin^2(\phi). 
\eeq 

Thus, in conclusion, this purely objective local formulation with 
erstwhile hidden variables $\theta, \gamma_{x}$ and $\gamma_{y}$ 
duplicates QM.


\begin{references}
\bibitem{spin} Bell actually considered spin-$1/2$ particles; but, as 
all agree that their description is isomorphic to that of `photons,' I 
restrict my discussion to radiation where visualization is cleaner.  
Spin demands special consideration discussed somewhat in \cite{KSB}.  
In any case, the argument presented herein, while crying for 
didactical elaboration of ancillary matters including spin, is  
rigorous and complete as it stands.
\end{references}
\end{document}